\documentclass[prd,aps,a4paper,superscriptaddress,twocolumn,nofootinbib]{revtex4-2}
\usepackage{graphicx}
\usepackage{color}
\usepackage{xcolor}
\usepackage{dcolumn}
\usepackage{bm}
\usepackage{slashed}
\usepackage{amsmath}
\usepackage{latexsym}
\usepackage{amssymb}
\usepackage{mathrsfs}
\usepackage{amsfonts}
\usepackage{xspace}
\usepackage{ulem}
\allowdisplaybreaks
\begin{document}
\title{The effect of the gravitational constant variation on the phase of gravitational waves}

\author{Jiachen An} 
\affiliation{Institute for Frontiers in Astronomy and Astrophysics, Beijing Normal University, Beijing 102206, China}
\affiliation{Department of Astronomy, Beijing Normal University, Beijing 100875, China}
\author{Bing Sun} 
\affiliation{Department of Basic Courses, Beijing University of Agriculture, Beijing 102206, China}
\author{Zhoujian Cao
\thanks{corresponding author}} \email[Zhoujian Cao: ]{zjcao@amt.ac.cn}
\affiliation{Institute for Frontiers in Astronomy and Astrophysics, Beijing Normal University, Beijing 102206, China}
\affiliation{Department of Astronomy, Beijing Normal University, Beijing 100875, China}
\affiliation{School of Fundamental Physics and Mathematical Sciences, Hangzhou Institute for Advanced Study, UCAS, Hangzhou 310024, China}

\begin{abstract}
We have previously investigated the effect of the gravitational constant variation on the gravitational wave propagation. Pure theoretical analysis indicates that the leading order effect of the gravitational constant variation corrects the amplitude of gravitational waves, and the second order effect corrects the phase of gravitational waves. As the matched filtering technique is used by gravitational wave data analysis, the phase is more important than the amplitude. In the current paper we use LIGO-VIRGO-KAGRA data to constrain the gravitational constant variation. Our findings indicate that we need to wait until the distance of the detected gravitational wave events and/or the signal-to-noise ratio increases by $2n$ orders compared to the current detection state, and then we can use phase correction to get constraint $|\frac{G'}{G}|<10^{-n}$/yr for gravitational wave events without electromagnetic counterparts. For gravitational wave events with electromagnetic counterparts which provide the information of source's luminosity distance, the phase correction can almost always be neglected to constrain $\frac{G'}{G}$.
\end{abstract}

\maketitle

\section{Introduction}
There are three primary reasons to question whether the gravitational constant varies with time and/or space. One is the large number hypothesis proposed by Dirac \cite{DIRAC1937}. The second one is the possible explanation of dark matter \cite{GOLDMAN1992219,2005JCAP...05..003M}, dark energy \cite{2007IJMPD..16.1791R} and even Hubble tension \cite{PhysRevD.104.L021303,2024arXiv241115301B}. The third one is that many alternative theories to general relativity, including extra dimension \cite{PhysRevLett.52.489}, can be effectively took as Einstein equation together with a varying gravitational constant \cite{WANG2022137416}. In other words, constraining the variation of the gravitational constant through observations can provide profound insights into fundamental physics, particularly gravity and cosmology.

People have paid much attention to constrain the gravitational constant variation through laboratory experiments and astrophysical observations. Laboratory experiments employ the Newtonian inverse-square law framework to measure variations in the gravitational constant relative to the distance between two objects \cite{PhysRevLett.108.081101,PhysRevLett.126.211101}. The astrophysical observations which have been used to constrain the gravitational constant variation include the solar evolution \cite{PhysRevLett.36.833}, lunar occultations and eclipses \cite{1973Natur.241..519M}, paleontological evidences \cite{PhysRevLett.72.454}, white dwarf cooling and pulsations \cite{2013JCAP...06..032C}, neutron star masses and ages \cite{PhysRevLett.77.1432}, star cluster evolutions \cite{1996A&A...312..345D}, big bang nucleosynthesis abundances \cite{2020EPJC...80..148A}, asteroseismology \cite{2019ApJ...887L...1B}, lunar laser ranging \cite{2018CQGra..35c5015H}, evolutions of planetary orbits \cite{2018NatCo...9..289G}, binary pulsars \cite{2019MNRAS.482.3249Z}, high-resolution quasar spectra \cite{2021GReGr..53...37L}, gravitational wave observations of binary neutron stars \cite{PhysRevLett.126.141104,WANG2022137416} and supernovae \cite{2018JCAP...10..052Z}.

Variations in the gravitational constant can influence the gravitational waveform \cite{PhysRevD.81.064018,WANG2022137416,An2023,SUN2024138350}. Consequently, people can use matched filtering technique to constrain the gravitational constant variation based on gravitational wave detection.

Phenomenologically, the observed gravitational constant is an effective gravitational constant \cite{PhysRevLett.108.081101,PhysRevLett.126.211101}. There are two ways to understand the effective gravitational constant. One is taking it as a dynamical field on the spacetime \cite{PhysRevD.81.064018,WANG2022137416}. The other one is taking it as a pre-fixed field without any dynamics \cite{An2023,SUN2024138350}. If the effective gravitational constant is a dynamical field, the dipole radiation of gravitational wave will appear \cite{PhysRevLett.108.081101,PhysRevLett.126.211101}. In contrast, if the effective gravitational constant is somehow a pre-fixed field, There is no more radiation of gravitational wave than that predicted by general relativity. But the variation of the effective gravitational constant does affect the propagation of gravitational waves \cite{PhysRevD.81.064018,WANG2022137416}. In fact when the effective gravitational constant is a dynamical field on the spacetime, the same effect on the propagation of gravitational waves will appear.

Along with our previous works \cite{An2023,SUN2024138350}, we still take the effective gravitational constant as a pre-fixed field in the current paper. In the next section we review the variation of the effective gravitational constant effects on the propagation of gravitational waves. Especially, the corrected gravitational waveform is explained there. After that we apply the corrected waveform to LIGO-VIRGO-KAGRA (LVK) data to constrain the gravitational constant variation. We will find that the amplitude correction of the signal dominates the data while the phase correction is undetectable for current detection accuracy. Following that, we use Fisher matrix analysis to predict in what situations the phase correction should be considered. Finally, we conclude our paper in the last section. We use geometric units $c=G=1$ through the current paper.

\section{gravitational waveform correction due to the variation of the effective gravitational constant}
In the previous works \cite{An2023,SUN2024138350}, we have found that both the amplitude and the phase of gravitational waves will be corrected as (Eq.~(66) of \cite{SUN2024138350})
\begin{align}
&\tilde{h}=\frac{1+\Xi_{\text{d}}/4\omega^2}{1+\Xi_{\text{s}}/4\omega^2}\sqrt{\frac{G_{\text{d}}}{G_{\text{s}}}}e^{-i(\int\Xi\,d\zeta)/2\omega}\tilde{h}_0,\\
&\Xi=\frac{3}{4}\left(\frac{G'}{G}\right)^2,\label{eq3}
\end{align}
where $G'$ means the derivative of the effective gravitational constant with respect to the affine parameter $\zeta$ of the trajectory when the gravitational wave propagates. In the above equation, $\omega$ means the circular frequency of the gravitational wave, subscripts ``$\text{d}$'' and ``$\text{s}$'' mean the detector position and the source position respectively, $\tilde{h}_0$ means the waveform predicted by standard general relativity (GR) in frequency domain, and $\tilde{h}$ means the corrected waveform in frequency domain. For compact binary coalescences (CBCs), we can equivalently write the waveform correction as
\begin{align}
&\tilde{h}(f)=\gamma(1+A(Mf)^{-2})e^{i\Omega(Mf)^{-1}}\tilde{h}_0(f),\label{eq1}\\
&\gamma\equiv\sqrt\frac{G_{\text{d}}}{G_{\text{s}}},\\
&A\equiv\frac{M^2(\Xi_{\text{d}}-\Xi_{\text{s}})}{16\pi^2},\\
&\Omega\equiv-\frac{M}{4\pi}\int\Xi\,d\zeta.
\end{align}
Here we have used $f$ to denote the frequency of the gravitational wave and $M$ to denote the total mass of the CBC system.

Based on the fact that $\frac{G'}{G}$ is a small quantity, we can approximate that
\begin{align}
&\int\Xi\,d\zeta\approx\Xi D_L,\\
&\Xi_{\text{d}}-\Xi_{\text{s}}\approx0,
\end{align}
where $D_L$ is the luminosity distance between the CBC and the detector. Therefore, Eq.~(\ref{eq1}) becomes
\begin{align}
\tilde{h}(f)\approx\sqrt{\frac{G_{\text{d}}}{G_{\text{s}}}}e^{-i\frac{\Xi D_L}{4\pi f}}\tilde{h}_0(f).\label{eq2}
\end{align}

The factor $\sqrt{\frac{G_{\text{d}}}{G_{\text{s}}}}$ in Eq.~(\ref{eq2}) corresponds to the amplitude correction which has been investigated in our previous works \cite{An2023,SUN2024138350}. Note that this amplitude correction is fully degenerate with the parameter $D_L$ during the gravitational wave data analysis, this correction is not detectable if $D_L$ is unknown in advance. This means that binary black hole events which have no electromagnetic counterpart can not be used to constrain the variation of the effective gravitational constant. Since
\begin{align}
&G_{\text{d}}\approx G_{\text{s}}+G'D_L,\\
&\frac{G_{\text{d}}}{G_{\text{s}}}\approx1+\frac{G'}{G}D_L,\\
&\sqrt{\frac{G_{\text{d}}}{G_{\text{s}}}}\approx1+\frac{1}{2}\frac{G'}{G}D_L,
\end{align}
the amplitude correction is of the order of $O\left(\frac{G'}{G}\right)$.

The factor $\frac{\Xi D_L}{4\pi f}$ in Eq.~(\ref{eq2}) corresponds to the phase correction. Reminding Eq.~(\ref{eq3}), we can see that the phase correction introduced by the variation of the gravitational constant is about $O\left(\left(\frac{G'}{G}\right)^2\right)$. If people take the effective gravitational constant as a dynamical field on the spacetime, an additional $-4$ PN phase correction will appear \cite{PhysRevD.81.064018,WANG2022137416}, which is of the order of $O\left(\frac{G'}{G}\right)$.

In a formal sense, the factor $\frac{\Xi D_L}{4\pi f}$ corresponds to $1$ PN correction. Adopting the notation which was used in the LVK paper on the test of GR \cite{2021arXiv211206861T}, we have
\begin{align}
&-\frac{3}{16}\left(\frac{G'}{G}\right)^2MD_L=\frac{3}{128\eta}\varphi_2\delta\varphi_2,\\
&\left|\frac{G'}{G}\right|=\sqrt{\left|-\frac{\varphi_2\delta\varphi_2}{8\eta MD_L}\right|},\\
&\varphi_2=\frac{20}{9}\left(\frac{743}{336}+\frac{11}{4}\eta\right),
\end{align}
where $\varphi_2$ is the $1$ PN phase coefficient of the standard GR waveform, and $\eta$ is the symmetric mass ratio of two components in the CBC. We follow the tradition of using $\delta\varphi_2$ to denote the deviation from GR. Based on GWTC-3, LVK has got constraint
\begin{align}
\left\lvert\delta\varphi_2\right\rvert\lesssim0.1.
\end{align}
The CBCs reported in GWTC-3 have $D_L\sim(100,10000)\,\text{Mpc}\sim(10^{16},10^{18})\,\text{sec}$ and $M\sim(10,100)\,M_\odot\sim(10^{-5},10^{-4})\,\text{sec}$ in geometric units. This means that we have a rough constraint
\begin{align}
    \left\lvert\frac{G'}{G}\right\rvert\lesssim10^{-7}/\text{sec}\sim1/\text{yr}.\label{eq4}
\end{align}

\section{constrain the variation of the gravitational constant with LVK data}
According to our previous research on GW170817 \cite{An2023,SUN2024138350}, we assume that $|D_L\frac{G'}{G}| < 20$, and an electromagnetic detection showed that for GW170817 $D_L\sim40\,\text{Mpc}$ \cite{2018ApJ...854L..31C}. Consequently, we take the prior distribution of $\frac{G'}{G}$ as a uniform distribution in the range $|\frac{G'}{G}| < 10^{-14}$/sec $\sim10^{-7}$/yr.

The LVK data we use is provided by the Gravitational Wave Open Science Center (GWOSC). In such data, the glitch appeared in the LIGO-Livingston detector has been removed. The data of noise power spectral density (PSD) which we use is the same as the LVK corresponding research.

Regarding theoretical waveform template for GR, we use SEOBNRv4\_ROM for binary black holes and IMRPhenomPv2\_NRTidal for binary neutron stars. When the variation of gravitational constant is considered, the correction described in the above section is implemented.

For sources with no electromagnetic counterpart, our previous works \cite{An2023,SUN2024138350} have not considered yet. In the current work we choose GW200129\_065458, which is the event with the highest signal-to-noise ratio in GWTC-3, as the example to do the analysis. We set the same prior distribution of source parameters as \cite{PhysRevX.13.041039}. The resulted posterior distribution of $\frac{G'}{G}$ is a uniform distribution as its prior distribution. This fact is consistent to the rough estimation (\ref{eq4}).

\begin{figure*}
\begin{tabular}{cc}
\includegraphics[width=0.45\textwidth]{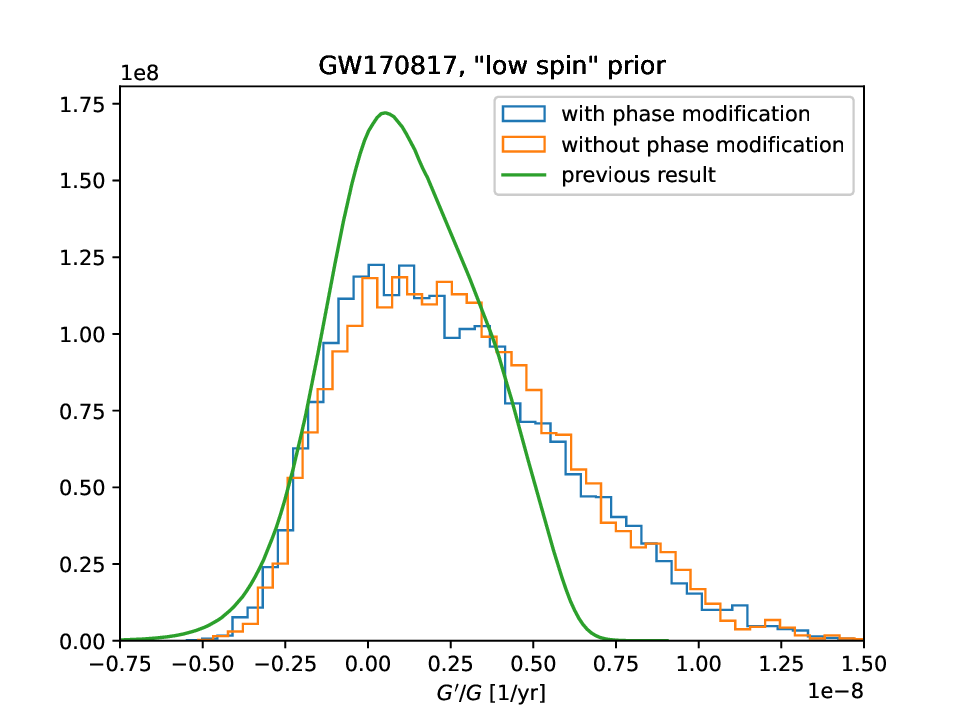}&
\includegraphics[width=0.45\textwidth]{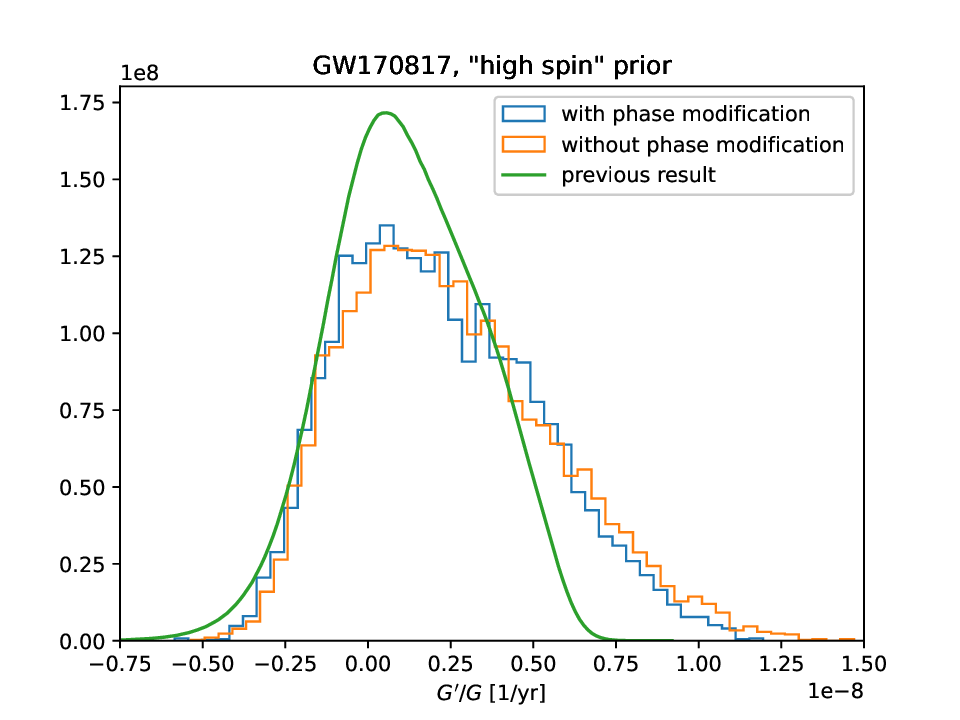}
\end{tabular}
\caption{Posterior distribution of $\frac{G'}{G}$ based on GW170817 data and waveform model IMRPhenomPv2\_NRTidal with the correction of variation of the gravitational constant. The green line corresponds to the result which we got in \cite{SUN2024138350}.}\label{fig1}
\end{figure*}

For sources with an electromagnetic counterpart, our previous works \cite{An2023,SUN2024138350} did not consider the phase correction. In the following, we check the improvement of constraint when the phase correction is taken into consideration. Regarding GW170817, we use the result of electromagnetic observations \cite{2017ApJ...848L..12A} as the prior distribution of the sky location. We also assume a Gaussian prior of luminosity distance $D_L$ according to the result of electromagnetic observations \cite{2018ApJ...854L..31C,2019PhRvX...9a1001A}. The priors of other parameters are set according to the result got in \cite{2019PhRvX...9a1001A}. Like \cite{2019PhRvX...9a1001A}, we also investigate both a ``low-spin'' prior study and a ``high-spin'' prior study. We plot the resulted posterior distribution of $\frac{G'}{G}$ in Fig.~\ref{fig1}. For comparison, we also plot the result we got in \cite{SUN2024138350}. In that work we did a simple estimation based on the measured luminosity $D_{L,\text{GW}}$ through gravitational waves and the luminosity $D_{L,\text{EM}}$ through optics. Different to the current work the gravitational wave data were not used at all there. We can see the consistency between these results.

Fig.~\ref{fig1} indicates that the phase correction negligibly improves the constraints on the variation of the gravitational constant. As we analyzed in the above section, the phase correction is of the order of $O\left(\left(\frac{G'}{G}\right)^2\right)$ instead of $O\left(\frac{G'}{G}\right)$ as the amplitude correction. Therefore, the result of Fig.~\ref{fig1} is consistent to our analysis. And this negligible improvement is also consistent to the uniform posterior result of binary black holes.

\section{Fisher matrix analysis on the phase correction of the variation of the gravitational constant}
In the above section we have applied the corrected waveform model due to the variation of the effective gravitational constant to the LVK data. Unfortunately we find that the phase correction is negligibly small for the current gravitational wave observation data. We expect that when the sensitivity of gravitational wave detectors is improved, the phase correction may be detectable. In this section, we use Fisher matrix analysis to check under what kind of situation the phase correction should be considered.

For a certain frequency domain waveform $\tilde{h}(f;\theta_1,\cdots,\theta_n)$ and a one-side noise power spectral density $S_N(f)$, the components of the Fisher information matrix can be written as
\begin{align}
\Gamma_{\theta_i\theta_j}=\left(\partial_{\theta_i}\tilde{h}|\partial_{\theta_j}\tilde{h}\right),
\end{align}
where the noise-weighted inner product is defined as
\begin{align}
\left(p|q\right)\equiv2\int_0^\infty\frac{p\bar{q}+\bar{p}q}{S_N}df,
\end{align}
where the over bar means the complex conjugate. Here we use $\theta_i$ to represent the gravitational waveform parameters.

Based on the waveform (\ref{eq2}), (\ref{eq3}) and noting that $\tilde{h}_0\propto\frac{1}{D_L}$, we have
\begin{align}
h(f;G'/G,D_L,\vec{\theta})=
e^{\frac{1}{2}\left(\frac{G'}{G}\right)D_L}
e^{-i\frac{\frac{3}{4}\left(\frac{G'}{G}\right)^2 D_L}{4\pi f}}
\frac{1}{D_L}
\zeta_0(f;\vec{\theta}),
\end{align}
Here we have denoted the parameters other than $G'/G$ and $D_L$ as $\vec{\theta}$, and $\zeta_0\equiv D_Lh_0$. Consequently, we have
\begin{align}
&\partial_{D_L}h=\left(\frac{G'}{2G}-\frac{1}{D_L}-i\frac{3}{16\pi f}(\frac{G'}{G})^2\right)h,\\
&\partial_{G'/G}h=\left(\frac{D_L}{2}-i\frac{3}{8\pi f}\frac{G'}{G}D_L\right)h,\\
&\partial_{\theta_k}h=\xi_k(f;\vec{\theta})h.
\end{align}

Accordingly, we have the Fisher information matrix with form
\begin{align}
\Gamma_{\mu\nu}&=\frac{e^{\left(\frac{G'}{G}\right)D_L}}{D_L^2}\gamma_{\mu\nu},\label{eq6}\\
\gamma_{(G'/G)(G'/G)}&=\int_0^\infty\left[D_L^2+4\left(\frac{G'}{G}D_L\frac{3}{8\pi f}\right)^2\right]\frac{
    \left|\zeta_0\right|^2}{{S_N}}df,\\
\gamma_{(G'/G)D_L}&=\int_0^\infty\left[-2+\frac{G'}{G}D_L\nonumber\right.\\
&\left.+ O\left(\left(\frac{G'}{G}\right)^3\right)\right]\frac{
        \left|\zeta_0\right|^2}{{S_N}}df,\\
\gamma_{(G'/G)\theta_k}&=\int_0^\infty\left[2\Re(\xi_k)D_L\nonumber\right.\\
&\left.- 4\Im(\xi_k)\frac{G'}{G}D_L\frac{3}{8\pi f}\right]\frac{
        \left|\zeta_0\right|^2}{{S_N}}df,\\
\gamma_{D_LD_L}&=\int_0^\infty\left[4\frac{1}{D_L^2}-4\frac{G'}{G}\frac{1}{D_L}\nonumber\right.\\
&\left.+ \left(\frac{G'}{G}\right)^2+O\left(\left(\frac{G'}{G}\right)^3\right)\right]\frac{\left|\zeta_0\right|^2}{{S_N}}df,\\
\gamma_{D_L\theta_k}&=\int_0^\infty\left[-4\Re(\xi_k)\frac{1}{D_L}+2\Re(\xi_k)\frac{G'}{G}\nonumber\right.\\
&\left.- 4\Im(\xi_k)\left(\frac{G'}{G}\right)^2\frac{3}{16\pi f}\right]\frac{\left|\zeta_0\right|^2}{{S_N}}df.
\end{align}
In the above equations, we have used indexes $\mu$ and $\nu$ to denote both parameter $\frac{G'}{G}$, $D_L$ and others, while we have used $k$ to denote only other parameters. Notations $\Re$ and $\Im$ mean taking the real part and the imaginary part respectively. We present the detail calculation in the Appendix, and the result is that the measurement accuracy
\begin{align}
\Delta\frac{G'}{G}\propto\frac{1}{e^{\frac{1}{2}\left(\frac{G'}{G}\right)D_L}|\frac{G'}{G}|}.\label{eq7}
\end{align}
Eq.~(\ref{eq2}) shows that the signal-to-noise ratio $\rho^2=(h|h)\propto\frac{e^{\left(\frac{G'}{G}\right)D_L}}{D_L^2}$. Consequently, (\ref{eq7}) becomes
\begin{align}
\Delta\frac{G'}{G}\propto\frac{1}{\rho D_L|\frac{G'}{G}|}.\label{eq8}
\end{align}

Our data analysis result for binary black hole done in the above section tells us that the current measurement accuracy is about $\Delta\frac{G'}{G}\sim10^{-7}$/sec. For standard GR, $G'=0$, which means that the measurement accuracy of $\frac{G'}{G}$ diverges. This is consistent to the fact that the Fisher matrix is singular for standard GR. But for real data, a variety of noise will result in a Gaussian distribution of $\frac{G'}{G}$ with the center deviating from the origin, and therefore the real measurement accuracy of $\frac{G'}{G}$ will not diverge. Considering such real data analysis fact, we can approximate that the mean value $\overline{(\frac{G'}{G})}\propto\Delta\frac{G'}{G}$. Then Eq.~(\ref{eq8}) becomes
\begin{align}
\Delta\frac{G'}{G}\propto\frac{1}{\sqrt{\rho D_L}}.
\end{align}
The above result means that, in order to make the measurement of the variation of the gravitational constant $\frac{G'}{G}$ reach an accuracy of the order of $10^{-7-n}$/sec $\sim10^{-n}$/yr, we need $\rho D_L$ to increase by $2n$ orders compared to the current state-of-the-art experiment.

If we have gotten the information of $D_L$ by methods which are independent of the gravitational wave detection, according to Eq.~(\ref{a3}) in the Appendix, we have
\begin{align}
    1/\sigma^2_{G'/G}&=\frac{e^{\left(\frac{G'}{G}\right)D_L}}{D_L^2}\left[\gamma_{(G'/G)(G'/G)}-\nonumber\right.\\
    &\left.\sum_{i,j}(D^{-1})_{ij}\gamma_{(G'/G)\theta_i}\gamma_{(G'/G)\theta_j}\right].\label{1/sigma_dlnG}
\end{align}
In Eq.~(\ref{1/sigma_dlnG}), $D_{ij}=\gamma_{ij}$ and
\begin{align}
    \gamma_{(G'/G)(G'/G)}&=\int_0^\infty\left[D_L^2+4\left(\frac{G'}{G}D_L\frac{3}{8\pi f}\right)^2\right]\frac{
        \left|\zeta_0\right|^2}{{S_N}}df,\\
        &=\int_0^\infty D_L^2\left[1+4\left(\frac{G'}{G}\frac{3}{8\pi f_{\rm GW}}\right)^2\right]\frac{
            \left|\zeta_0\right|^2}{{S_N}}df,\label{eq5.1}
\end{align}
\begin{align}
    \gamma_{(G'/G)\theta_k}&=\int_0^\infty\left[2\Re(\xi_k)D_L\nonumber\right.\\
&\left.- 4\Im(\xi_k)\frac{G'}{G}D_L\frac{3}{8\pi f}\right]\frac{
        \left|\zeta_0\right|^2}{{S_N}}df,\\
        &=\int_0^\infty D_L\left[2\Re(\xi_k)\nonumber\right.\\
&\left.- 4\Im(\xi_k)\left(\frac{G'}{G}\frac{3}{8\pi f_{\rm GW}}\right)\right]\frac{
        \left|\zeta_0\right|^2}{{S_N}}df,\label{eq5.2}
\end{align}
where $f_{\rm GW}$ is the characteristic frequency of the detected gravitational wave. For LVK data we have $f_{\rm GW}\sim100$ Hz. Current measurement accuracy of $\frac{G'}{G}$, which is solely based on the amplitude correction, is about $\Delta\frac{G'}{G} < 10^{-14}$/s $\sim10^{-7}$/yr \cite{An2023,SUN2024138350}. Therefore, $\frac{G'}{G}\frac{3}{8\pi f_{\rm GW}}$, which is about $10^{-17}$, in Eq.~(\ref{eq5.1}) and Eq.~(\ref{eq5.2}) can be safely neglected. This can explain why we cannot see the improvement by the phase correction in Fig.~\ref{fig1}. Based on current measurement accuracy $\Delta\frac{G'}{G}$ $\sim10^{-7}$/yr, only if $f_{\rm GW}<10^{-16}$ Hz do we need to consider the phase correction; otherwise we can safely ignore the phase correction in the gravitational wave data analysis.

\section{Conclusion and discussion}
The nature of the gravitational constant is very important to fundamental physics. In standard general relativity theory, the gravitational constant is the coupling constant between the gravitational field and matter, while the self-interaction of the gravitational field is independent of the gravitational constant. If the gravitational constant varies, this situation will change. We have investigated the effect of the variation of the gravitational constant on the gravitational waveform before \cite{An2023,SUN2024138350}.

If one take the varying gravitational constant as a pre-fixed field on the spacetime, the variation of the gravitational constant will affect both the amplitude and the phase of the gravitational waveform. The amplitude correction is of the order of $O(\frac{G'}{G})$ and the phase correction is of the order of $O((\frac{G'}{G})^2)$. Since the gravitational wave detection is more sensitive to the phase than to the amplitude, it is still interesting to investigate the effect of phase correction based on LVK data, even if the phase correction is second order with respect to the variation of $G$. We have done such investigation in the current paper.

Since the amplitude correction completely is degenerate with the gravitational wave source luminosity, the amplitude correction alone cannot be used to constrain the variation of the gravitational constant. After we take the phase correction into consideration, such degeneracy is broken. Based on such consideration, we firstly use binary black hole gravitational wave event to constrain the variation of the gravitational constant. We find that the resulted constraint is looser than $|\frac{G'}{G}| < 10^{-14}$/sec $\sim10^{-7}$/yr.

We also investigate GW170817 which has an electromagnetic counterpart. The information about luminosity is taken as the prior distribution. Then we get roughly the same posterior distribution with and without phase correction, and the resulted constraint of the variation of the gravitational constant is consistent to the result which we got in \cite{An2023,SUN2024138350}.

Our analysis of LVK data suggests that the phase correction negligibly improves the constraints on the variation of the gravitational constant. One can understand that this fact is due to the relatively low signal-to-noise ratio of the gravitational wave events. Then it is interesting to investigate when we should take the phase correction into consideration. We use Fisher information matrix analysis to do so. We find that in order to make the measurement of the variation of the gravitational constant $\frac{G'}{G}$ reach an accuracy of the order of $10^{-7-n}$/sec $\sim10^{-n}$/yr, we need $\rho D_L$ to increase by $2n$ orders compared to the current state-of-the-art experiment. For events where the luminosity is known, such as those with electromagnetic counterparts, only if $f_{\rm GW}<10^{-16}$ Hz do we need to consider the phase correction.

A final caution is that the conclusions presented in this paper are based on the assumption that the varying gravitational constant is a pre-fixed field on spacetime. If another viewpoint about effective gravitational constant is taken, for example, as stated in \cite{PhysRevD.81.064018,WANG2022137416}, the effect of the varying gravitational constant during the stage of the gravitation wave generation will be important, and the constraint result will be different.

\section*{Acknowledgments}
This research has made use of data or software obtained from the Gravitational Wave Open Science Center (gwosc.org), a service of LIGO Laboratory, the LIGO Scientific Collaboration, the Virgo Collaboration, and KAGRA. LIGO Laboratory and Advanced LIGO are funded by the United States National Science Foundation (NSF) as well as the Science and Technology Facilities Council (STFC) of the United Kingdom, the Max-Planck-Society (MPS), and the State of Niedersachsen/Germany for support of the construction of Advanced LIGO and construction and operation of the GEO600 detector. Additional support for Advanced LIGO was provided by the Australian Research Council. Virgo is funded, through the European Gravitational Observatory (EGO), by the French Centre National de Recherche Scientifique (CNRS), the Italian Istituto Nazionale di Fisica Nucleare (INFN) and the Dutch Nikhef, with contributions by institutions from Belgium, Germany, Greece, Hungary, Ireland, Japan, Monaco, Poland, Portugal, Spain. KAGRA is supported by Ministry of Education, Culture, Sports, Science and Technology (MEXT), Japan Society for the Promotion of Science (JSPS) in Japan; National Research Foundation (NRF) and Ministry of Science and ICT (MSIT) in Korea; Academia Sinica (AS) and National Science and Technology Council (NSTC) in Taiwan.

This work was supported in part by the National Key Research and Development Program of China Grant No. 2021YFC2203001 and in part by the NSFC (No.~12475046, No.~12375046, No.~12021003 and No.~12005016). B. Sun is supported by Beijing University of Agriculture (QJKC-2023032). Z. Cao was supported by ``the Fundamental Research Funds for the Central Universities".

\appendix
\section{Inverse of the Fisher information matrix (\ref{eq6})}

In this appendix we calculate the inverse of the Fisher information matrix (\ref{eq6}). For a blocked matrix
\begin{align}
\gamma=\left[\begin{matrix}
    A&B\\C&D
\end{matrix}\right],
\end{align}
where $A$ and $D$ are $m\times m$ and $n\times n$ matrices. If $D$ is invertible we have the inverse of $M$ as
\begin{align}
\gamma^{-1}&=\left[\begin{matrix}
    E&F\\G&H
\end{matrix}\right],\\
    E&=(A-BD^{-1}C)^{-1},\label{a3}\\
    F&=-(A-BD^{-1}C)^{-1}BD^{-1},\\
    G&=-D^{-1}C(A-BD^{-1}C)^{-1},\\
    H&=D^{-1}+D^{-1}C(A-BD^{-1}C)^{-1}BD^{-1}.
\end{align}

We let $A$ to denote the $G'/G$ and $D_L$ $2\times2$ part. And more we denote $\Theta_1=G'/G$, $\Theta_2=D_L$, then according to Eq.~(\ref{a3}), we have
\begin{align}
(E^{-1})_{ij}=\gamma_{\Theta_i\Theta_j}+\sum_{k,l}(D^{-1})_{kl}\gamma_{\Theta_i\theta_k}\gamma_{\Theta_j\theta_l},
\end{align}
Here we have used indexes $i,j=1,2$ to denote parameters $\frac{G'}{G}$, $D_L$ and $k$ and $l$ to denote other parameters.

We use superscript $(0)$, $(1)$ and $(2)$ to denote $O\left(1\right)$ part, $O\left(\frac{G'}{G}\right)$ part and $O\left(\left(\frac{G'}{G}\right)^2\right)$ part respectively,
\begin{align}
    (E^{-1})_{ij}^{(0)}&=\gamma_{\Theta_i\Theta_j}^{(0)}+\sum_{k,l}(D^{-1})_{kl}\gamma_{\Theta_i\theta_k}^{(0)}\gamma_{\Theta_j\theta_l}^{(0)},\\
    (E^{-1})_{ij}^{(1)}&=\gamma_{\Theta_i\Theta_j}^{(1)}+\sum_{k,l}(D^{-1})_{kl}\times\nonumber\\
    &\left(\gamma_{\Theta_i\theta_k}^{(0)}\gamma_{\Theta_j\theta_l}^{(1)}+\gamma_{\Theta_i\theta_k}^{(1)}\gamma_{\Theta_j\theta_l}^{(0)}\right),\\
    (E^{-1})_{ij}^{(2)}&=\gamma_{\Theta_i\Theta_j}^{(2)}+\sum_{k,l}(D^{-1})_{kl}\left(\gamma_{\Theta_i\theta_k}^{(0)}\gamma_{\Theta_j\theta_l}^{(2)}+\right.\nonumber\\
    &\left.\gamma_{\Theta_i\theta_k}^{(1)}\gamma_{\Theta_j\theta_l}^{(1)}+\gamma_{\Theta_i\theta_k}^{(2)}\gamma_{\Theta_j\theta_l}^{(0)}\right),\\
    \left|E^{-1}\right|^{(1)}&=(E^{-1})_{11}^{(0)}(E^{-1})_{22}^{(1)}+(E^{-1})_{11}^{(1)}(E^{-1})_{22}^{(0)}\nonumber\\
    &-2(E^{-1})_{12}^{(0)}(E^{-1})_{12}^{(1)},\\
    \left|E^{-1}\right|^{(0)}&=(E^{-1})_{11}^{(0)}(E^{-1})_{22}^{(0)}-{(E^{-1})_{12}^{(0)}}^2,\\
    \left|E^{-1}\right|^{(2)}&=(E^{-1})_{11}^{(0)}(E^{-1})_{22}^{(2)}+(E^{-1})_{11}^{(1)}(E^{-1})_{22}^{(1)}\nonumber\\
    &+(E^{-1})_{11}^{(2)}(E^{-1})_{22}^{(0)}-2(E^{-1})_{12}^{(0)}(E^{-1})_{12}^{(2)}\nonumber\\
    &-{(E^{-1})_{12}^{(1)}}^{2},\\
    (E^{-1})^{(0)}&=(E^{-1})_{11}^{(0)}
    \begin{bmatrix}
    1&-\frac{2}{D_L^2}\\
    -\frac{2}{D_L^2}&\frac{4}{D_L^4}\\
    \end{bmatrix},\\
    (E^{-1})_{11}^{(0)}&=D_L^2\left(\int_0^\infty\frac{\left|\zeta_0\right|^2}{{S_N}}df+\sum_{i,j}(D^{-1})_{ij}\times\right.\nonumber\\
    &\left.\int_0^\infty[2\Re(\xi_i)]\frac{\left|\zeta_0\right|^2}{{S_N}}df\int_0^\infty[2\Re(\xi_j)]\frac{\left|\zeta_0\right|^2}{{S_N}}df\right),\\
        (E^{-1})^{(1)}&=(E^{-1})_{11}^{(0)}
    \begin{bmatrix}
        0&\frac{G'}{G}\frac{1}{D_L}\\
        \frac{G'}{G}\frac{1}{D_L}&-4\frac{G'}{G}\frac{1}{D_L^3}
    \end{bmatrix}\nonumber\\
    &+(E^{-1})_{11}^{(1)}
    \begin{bmatrix}
        1&-\frac{1}{D_L^2}\\
        -\frac{1}{D_L^2}&0
    \end{bmatrix},\\
    (E^{-1})_{11}^{(1)}&=2\frac{G'}{G}D_L^2\sum_{i,j}(D^{-1})_{ij}\int_0^\infty[2\Re(\xi_i)]\frac{
        \left|\zeta_0\right|^2}{{S_N}}df\times\nonumber\\
        &\int_0^\infty[-4\Im(\xi_j)\frac{3}{8\pi f}]\frac{
    \left|\zeta_0\right|^2}{{S_N}}df  ,\\
        (E^{-1})^{(2)}&=(E^{-1})_{11}^{(0)}
    \begin{bmatrix}
        0&0\\
        0&(\frac{G'}{G})^2\frac{1}{D_L^2}
    \end{bmatrix}\nonumber\\
    &+(E^{-1})_{11}^{(1)}
    \begin{bmatrix}
        0&\frac{3}{4}\frac{G'}{G}\frac{1}{D_L}\\
        \frac{3}{4}\frac{G'}{G}\frac{1}{D_L}&-\frac{G'}{G}\frac{1}{D_L^3}
    \end{bmatrix}\nonumber\\
    &+(E^{-1})_{11}^{(2)}
    \begin{bmatrix}
        1&0\\
        0&0
    \end{bmatrix}  ,\\
    (E^{-1})_{11}^{(2)}&=(\frac{G'}{G})^2D_L^2\left[\int_0^\infty[4(\frac{3}{8\pi f})^2]\frac{
    \left|\zeta_0\right|^2}{{S_N}}df\right.\nonumber\\
    &+\sum_{i,j}(D^{-1})_{ij}\int_0^\infty[-4\Im(\xi_i)\frac{3}{8\pi f}]\frac{
    \left|\zeta_0\right|^2}{{S_N}}df\times\nonumber\\
    &\left.\int_0^\infty[-4\Im(\xi_j)\frac{3}{8\pi f}]\frac{
    \left|\zeta_0\right|^2}{{S_N}}df\right],\\
    \left|E^{-1}\right|^{(0)}&=0,\\
    \left|E^{-1}\right|^{(1)}&=0,\\
    \left|E^{-1}\right|^{(2)}&=\frac{1}{D_L^4}[4(E^{-1})_{11}^{(0)}(E^{-1})_{11}^{(2)}-{(E^{-1})_{11}^{(1)}}^2]\\
    &\propto(\frac{G'}{G})^2,\\
    (E^{-1})_{22}^{(0)}&\propto\frac{1}{D_L^2}.
\end{align}
According to (\ref{eq6}), $(\Delta\frac{G'}{G})^2=(\Gamma^{-1})_{11}=\frac{D_L^2}{e^{\left(\frac{G'}{G}\right)D_L}}(\gamma^{-1})_{11}$, and $(\gamma^{-1})_{11}=E_{11}=\frac{(E^{-1})_{22}}{\left|E^{-1}\right|}$, therefore
\begin{equation}
    (\Delta\frac{G'}{G})^2\propto\frac{1}{e^{\left(\frac{G'}{G}\right)D_L}(\frac{G'}{G})^2}.
\end{equation}

\bibliography{refs}

\end{document}